\documentclass[prb,aps,12pt,nofootinbib]{revtex4}
\usepackage{verbatim}
\usepackage{graphicx}
\usepackage{feynmp}
\usepackage{amsmath,amsfonts,amsthm,amssymb}
\newcommand{\be}{\begin{eqnarray}}
\newcommand{\ee}{\end{eqnarray}}

\begin{document}

\title{Kinetic Equations Governing Smoluchowski Dynamics in Equilibrium}
\author{Gene F. Mazenko, David D. McCowan and Paul Spyridis}
\affiliation{The James Franck Institute and the Department of Physics, The University of Chicago,
Chicago, Illinois 60637, USA}
\date{April 3, 2012}

\begin{abstract}

We continue our study of the statistical properties of particles in equilibrium obeying Smoluchowski dynamics. We show that the system is governed by a kinetic equation of the memory function form and that the memory function is given by one of the self-energies available via perturbation theory as introduced in previous work. We determine the memory function explicitly to second-order in an expansion in a pseudo-potential. The method we use allows for a straightforward computation of corrections via a formal expansion and we therefore view it as an improvement over the conventional mode-coupling theory (MCT) formalism where it is not clear how to make systematic corrections. In addition, the formalism we have introduced is flexible enough to allow for a wide array of different approximation schemes, including density expansions. The convergence criteria for our formal series are not worked out here, but the second order equation that we derive is promising in the sense that it leads to analytic and numerical results consistent with expectations from computer simulations of the hard sphere system in addition to replicating the desired features from conventional MCT (e.g., a two-step decay). These particular solutions will be discussed in forthcoming work.

\end{abstract}

\maketitle

\section{Introduction}

A powerful approach for studying the dynamics of systems of classical particles was presented in Ref. \onlinecite{FTSPD} (referred to here as FTSPD). In Ref. \onlinecite{SDENE} (referred to here as SDENE) this method was developed to study fluctuations in equilibrium for systems obeying Smoluchowski dynamics (SD)\cite{SD}. A parallel development is carried out in Ref. \onlinecite{Newtonian} for systems obeying Newtonian dynamics. The treatment of SD is extended here to include a full self-consistent\cite{self-const} treatment of the separation of statics and dynamics and a derivation of a kinetic equation -- valid to second order in perturbation theory -- governing the density fluctuations over the entire time range.

In SDENE, the fluctuation kinetics are described in terms of a self-energy structure.  These self-energies can be conveniently obtained in a perturbation theory expansion in a pseudo-potential. Furthermore, the self-energies are divided into single-particle and collective contributions.  In SDENE, we focused on the collective contribution which governs the long-time slow kinetics in the problem, and showed that one can find a simple self-consistent relationship between the static structure factor and the zero-frequency component of the collective part of the self-energies.

In this paper, we fully analyze the single-particle contributions to the self-energies. We show the following:

\begin{enumerate}
\item
The single-particle contribution to the self-energy can be associated with the equation of state governing the system.

\item
While the collective degrees of freedom dominate the long-time dynamics in this system, the single-particle degrees of freedom govern the early-time kinetics and the approach to the slow-regime.

\item
We derive here the kinetic equation of the memory-function type valid at second-order in perturbation theory and including both single-particle and collective contributions.
\end{enumerate}

It is our intention to solve this kinetic equation numerically in future work\cite{KE_Dave}, and to show that the analysis of the collective contribution in SDENE can be extended analytically to obtain a two-step kinetic process similar to that obtained from mode coupling theory (MCT)\cite{KE_Paul}.

MCT represents the current \textit{de facto} theoretic description of dense fluids and the transition from fluid to glassy state\cite{Crisanti, Goetze, Das}.  However, MCT is limited by its ad hoc construction and lacks a mechanism to institute systematic corrections. We derive here, on the other hand, a form of the kinetic equation of the memory-function type used in MCT, and our theory provides the crucial advantage of well-defined, perturbative corrections.  Thus, one can use our methods to study vertex corrections, three- and higher-mode corrections to the standard MCT two-mode form, and high-frequency effects.

We begin with a brief review of the previous work to establish context and notation. We next develop the equation of state described above and then complete the second-order vertex function. In the final section, we derive the kinetic equation and discuss the memory function at its heart.

\section{Review of previous work}
The key components of the field theory approach are the two-point matrix cumulant functions, $G_{ij}(q,\omega)$, and the two-point irreducible vertex functions, $\Gamma_{ij}(q,\omega)$, with $i$ and $j$ running over the fields $\rho$ and $B$ where $\rho$ is the particle density and $B$ is a response field. The kinetic equation of interest results from an analysis of Dyson's equation which takes the form
\be
\sum_{k}\Gamma_{ik}G_{kj} = \delta_{ij}.
\label{eq:1}
\ee

In SDENE it was shown that the two-point irreducible vertex can be separated into two contributions,
\be
\Gamma_{ij} = \gamma_{ij} + K_{ij},
\label{eq:2}
\ee
where $\gamma_{ij}$ is the single-particle contribution and $K_{ij}$ is the collective contribution. The second-order contribution, $K_{ij}^{(2)}$, was derived and partially analyzed. It was shown that $K_{ij}^{(2)}$ itself satisfies a fluctuation-dissipation relation while remaining a quadratic functional of the two-point matrix correlation function, $G_{ij}$. The single-particle contribution to the two-point vertex, $\gamma_{ij}$, is defined as the inverse of the quantity
\be
\mathcal{G}_{ij}=\textrm{Tr}~\phi_{i}\phi_{j}e^{H\cdot\phi+\Delta W}
\label{eq:3}
\ee
such that
\be
\sum_{k}\gamma_{ik}\mathcal{G}_{kj} = \delta_{ij}
\label{eq:4}
\ee
where the fields $\Phi = (\rho, B)$ are one-particle additive,
\be
\Phi_i = \sum_{\alpha=1}^N \phi^{\alpha}_i,
\ee
$H=(H_{\phi},H_{B})$ is a conjugate external coupling field, and the term $\Delta W$ contains the pseudo-potential interaction.

The form of the interaction $\Delta W$ is defined carefully in FTSPD, but to second order in the pseudo-potential is given by
\be
\Delta W &=& \Delta W^{(1)} + \Delta W^{(2)} + ...
\label{eq:6}
\ee
where
\be
\Delta W^{(1)} &=& \sum_u F_u G_u,\\
\Delta W^{(2)} &=& \frac{1}{2} \sum_{u,v} F_u F_v G_{uv},
\ee
and
\be
F_i = \sum_j \sigma_{ij}\phi_j,
\ee
and where the interaction matrix $\sigma_{ij}$ is defined by
\be
\sigma_{ij}(q)=V(q)(\delta_{i\rho}\delta_{jB} +\delta_{iB}\delta_{j\rho})
\ee
where $V(q)$ is the Fourier transform of the potential. (We have not found it confusing to use the same symbol for both the coordinate- and wavenumber-space representations.)

The trace, $\textrm{Tr}$, is defined carefully in FTSPD and SDENE, however we do not need the details in defining Eq.(\ref{eq:3}) since \textit{all} the noninteracting cumulants among the fields $\rho$ and $B$ are available for the non-interacting case in FTSPD. This suffices to determine $\mathcal{G}_{ij\ldots k}$.
We gave the solution for the noninteracting $\gamma_{ij}^{(0)}$ in both FTSPD and SDENE and provide a summary of the results in Appendix \ref{app:Vertices}.

In general, the procedure is to determine $\gamma_{ij}$ and $K_{ij}$, then solve Eq.(\ref{eq:1}) to obtain the two-point correlation functions $G_{\rho\rho}$, $G_{\rho B}$, and $G_{B\rho}$. In the special -- but very important -- case where the system is in thermal equilibrium, one finds that there is a simple fluctuation-dissipation relation (FDR) between the two-point quantities,
\be
G_{\rho B}(q,\omega)-G_{B\rho} (q,\omega) = i\beta\omega G_{\rho\rho}(q,\omega).
\ee

In SDENE, we showed that one has in this case a simple kinetic equation satisfied by the density-density correlation function
\be
\frac{\partial}{\partial t}G_{\rho\rho}(q,t) = -\bar{D}q^2\bar{\rho}S^{-1}(q)G_{\rho\rho}(q,t)
-\bar{D}q^2\int_{t'}^t ds \beta^2\bar{\rho}\Sigma_{BB}(q,t-s)\frac{\partial}{\partial s}G_{\rho\rho}(s)
\ee
where $\Sigma_{BB}(q,t)$ is the kinetic contribution to $\Gamma_{BB}$ occurring in Eq.(\ref{eq:1}), $S(q)$ is the static structure factor, $\bar{\rho}$ is the average density, and $\bar{D} =k_{B}TD$ is the product of the temperature and diffusion coefficient. In the same work, we determined the collective contribution to $\Gamma_{ij}$ to second-order in an expansion in a pseudo-potential and established that $\Gamma_{B\rho}^{(2)}$ and $\Gamma_{BB}^{(2)}$ themselves satisfy a FDR and are quadratic functionals of the exact density-density correlation function $G_{\rho\rho}(q,t)$. Here we want to determine the ``single-particle" contribution to $\Gamma_{BB}$.

\section{Equation of state}

In our approach here, we generate approximations for both the static and dynamic properties. In SDENE, we showed how approximations for the static structure factor entered the analysis and we used the collective part of the self-energy, $K_{ij}$, to make contact with the equilibrium statics via the static structure factor. Here, we want to show how the equation of state enters the development.

\subsection{Equation of state}
In FTSPD, we established the fundamental identity for the one point quantity
\be
G_i = \textrm{Tr}~\phi_i e^{H\cdot\phi + \Delta W}
\ee
where $i$ labels space, time, and fields $\rho$ or $B$. This is the equation of state. For zero external field, $H=0$, and keeping terms to second-order in the pseudo-potential as given by Eq.(\ref{eq:6}), we find the one-point quantity
\be
G_i &=& \textrm{Tr}~\phi_i\bigg[1 + \Delta W^{(1)}
+ \Delta W^{(2)}+\frac{1}{2}(\Delta W^{(1)})^2\bigg] + ...
\nonumber\\
&=& \textrm{Tr}~\phi_i + \textrm{Tr}~\phi_i\sum_u F_u G_u
+ \textrm{Tr}~\phi_i\frac{1}{2} \sum_{u,v} F_u F_v (G_{uv}+G_uG_v) + ...
\nonumber\\
&=& \textrm{Tr}~\phi_i + \sum_{u,k}\textrm{Tr}~\phi_i\sigma_{uk}\phi_k G_u
+ \frac{1}{2} \sum_{u,v,k,\ell} \textrm{Tr}~\phi_i \sigma_{uk}\phi_k\sigma_{v\ell}\phi_{\ell}(G_{uv}+G_uG_v) + ...
\nonumber\\
&=& G_i^{(0)} + \sum_{u,k}G_{ik}^{(0)}\sigma_{ku} G_u
+ \frac{1}{2} \sum_{u,v,k,\ell} G_{ik\ell}^{(0)} \sigma_{ku}\sigma_{\ell v}(G_{uv}+G_uG_v) + ...
\label{eq:13}
\ee

Using wavenumber and frequency labels such that $1=(k_1,\omega_1)$, we have the zeroth-order contribution (valid for uniform systems)
\be
G_{i}^{(0)}(q_1,\omega_1)= \delta_{i\rho} \delta(1) \rho_0
\ee
where we introduce the notation $\delta(1) = (2\pi)^d\delta(q_1)2\pi\delta(\omega_1)$ and where $\rho_0$ is the density in the absence of interactions in the grand canonical ensemble.

Next, we have the first-order contribution
\be
G_{i}^{(1)}(1)=G_{ij}^{(0)}(1\bar{2})\sigma_{jk}(\bar{2}\bar{3})G_{k}(\bar{3})
\ee
where we now move to a convention where summation over repeated indices and integration over repeated, barred variables are implied. At all orders, the two point cumulant has the form
\be
G_{ij}(12)=G_{ij}(1)\delta(1+2)
\ee
(with $\delta(1+2) = (2\pi)^d\delta(q_1+q_2)2\pi\delta(\omega_1+\omega_2)$) due to translational invariance, and likewise the full one-point is
\be
G_{i}(1)&=&\delta_{i\rho}\delta(1)\bar{\rho}
\ee
where $\bar{\rho}$ is the average density.

It is easy to show that the first-order contribution then yields
\be
G_{i}^{(1)}(1)&=&G_{iB}^{(0)}(1\bar{2})\sigma_{B\rho}(\bar{2}\bar{3})G_{\rho}(\bar{3})
\nonumber\\
&=&G_{iB}^{(0)}(10)V(0)G_{\rho}(0)
\nonumber\\
&=&(-\beta G_{i}^{(0)}(1))V(0)\bar{\rho}
\nonumber\\
&=&-\beta \bar{\rho}V(0)\rho_0\delta(1)\delta_{i\rho}
\label{eq:18}
\ee
where, in the next to last line, we used the identity $G_{\rho B}^{(0)}(10)=-\beta G_{\rho}(1)$ discussed in Appendix \ref{app:ID}.

Turning to the second-order contribution, we have
\be
G_{i}^{(2)}(1)= G_i^{(2,1)}(1)+G_i^{(2,2)}(1)
\ee
where
\be
G_{i}^{(2,1)}(1)=
\frac{1}{2}G_{ijk}^{(0)}(1\bar{2}\bar{3})\sigma_{ju}(\bar{2}\bar{4})
\sigma_{kv}(\bar{3}\bar{5})G_{u}(\bar{4})G_{v}(\bar{5})
\ee
and
\be
G_{i}^{(2,2)}(1)=
\frac{1}{2}G_{ijk}^{(0)}(1\bar{2}\bar{3})\sigma_{ju}(\bar{2}\bar{4})\sigma_{kv}(\bar{3}\bar{5})G_{uv}(\bar{4}\bar{5}).
\ee
We can express the three-point cumulants in terms of the three-point
irreducible vertex,
\be
G_{ijk}^{(0)}(123) = -G_{ix}^{(0)}(1\bar{4})G_{jy}^{(0)}(2\bar{5})G_{kz}^{(0)}(3\bar{6})\gamma_{xyz}^{(0)}(\bar{4}\bar{5}\bar{6}).
\label{eq:22}
\ee
(The non-interacting three-point vertex functions $\gamma_{ijk}^{(0)}$ are summarized in Appendix \ref{app:Vertices}.)
This gives,
\be
G_{i}^{(2,1)}(1)&=&
-\frac{1}{2}G_{ix}^{(0)}(1\bar{6})G_{jy}^{(0)}(\bar{2}\bar{7})G_{kz}^{(0)}(\bar{3}\bar{8})
\gamma_{xyz}^{(0)}(\bar{6}\bar{7}\bar{8})
\sigma_{ju}(\bar{2}\bar{4})\sigma_{kv}(\bar{3}\bar{5})G_u(\bar{4})G_v(\bar{5})
\nonumber\\
&=&-\frac{1}{2}G_{ix}^{(0)}(1\bar{6})\gamma_{xyz}^{(0)}(\bar{6}\bar{7}\bar{8})
G^{(1)}_y(\bar{7})G^{(1)}_z(\bar{8})
\nonumber\\
&=&-\frac{1}{2}G_{ix}^{(0)}(1\bar{6})\gamma_{xyz}^{(0)}(\bar{6}\bar{7}\bar{8})
(-\beta V(0)\bar{\rho})G^{(0)}_y(\bar{7})
(-\beta V(0)\bar{\rho})G^{(0)}_z(\bar{8})
\nonumber\\
&=&-\frac{1}{2}G_{ix}^{(0)}(1\bar{6})(\beta V(0)\bar{\rho})^2
(-\gamma_{xy}^{(0)}(\bar{6}\bar{7})G_y^{(0)}(\bar{7}))
\nonumber\\
&=&\frac{1}{2}(\beta V(0)\bar{\rho})^2G_i^{(0)}(1)
\nonumber\\
&=&\frac{1}{2}(\beta V(0)\bar{\rho})^2\rho_0\delta_{i\rho}\delta(1)
\label{eq:23}
\ee
where (in the fourth line) we have used the identity (established in Appendix \ref{app:ID})
\be
\gamma_{xyz}^{(0)}(12\bar{3})G_z^{(0)}(\bar{3})=-\gamma_{xy}^{(0)}(12).
\ee

Next, we have the more complicated second-order contribution
\be
G_{i}^{(2,2)}(1)=
\frac{1}{2}G_{ijk}^{(0)}(1\bar{2}\bar{3})\sigma_{ju}(\bar{2}\bar{4})\sigma_{kv}(\bar{3}\bar{5})G_{uv}(\bar{4}\bar{5})
\nonumber
\ee
\be
=-\frac{1}{2}G_{ix}^{(0)}(1\bar{2})
\gamma_{xyz}^{(0)}(\bar{2}\bar{3}\bar{4})
\tilde{G}_{yz}(\bar{3}\bar{4})
\label{eq:25}
\ee
where we again have an effective propagator
\be
\tilde{G}_{yz}(12)
=G_{y\ell}^{(0)}(1\bar{3})\sigma_{\ell p}(\bar{3}\bar{4})
G_{pu}(\bar{4}\bar{5})
\sigma_{uv}(\bar{5}\bar{6})
G_{vz}^{(0)}(\bar{6}2)
=\tilde{G}_{yz}(1)\delta(1+2).
\label{eq:29}
\ee

Looking at the $i=B$ component, we have
\be
G_{B}^{(2,2)}(1)&=&
-\frac{1}{2}G_{Bx}^{(0)}(1\bar{2})
\gamma_{xyz}^{(0)}(\bar{2}\bar{3}\bar{4})
\tilde{G}_{yz}(\bar{3}\bar{4})
\nonumber\\
&=&-\frac{1}{2}G_{B\rho}^{(0)}(1\bar{2})
\gamma_{\rho yz}^{(0)}(\bar{2}\bar{3}\bar{4})
\tilde{G}_{yz}(\bar{3}\bar{4}).
\ee
Enforcing the $\delta$-functions, we find
\be
G_{B}^{(2,2)}(1)&=&
-\frac{1}{2}\delta (1)G_{B\rho}^{(0)}(1,0)
\int d3 \gamma_{\rho yz}^{(0)}(0,3,-3)
\tilde{G}_{yz}(3)
\nonumber\\
&=&-\frac{1}{2}\delta (1)G_{B\rho}^{(0)}(1,0)
\nonumber\\
&&\times\int d3 \bigg[\gamma_{\rho \rho B}^{(0)}(0,3,-3)
\tilde{G}_{\rho B}(3)
+ \gamma_{\rho B\rho}^{(0)}(0,3,-3)
\tilde{G}_{B\rho }(3)\bigg].
\ee

Consulting the forms given in Appendix \ref{app:Vertices} for the three-point vertex functions, we see that each contributes at most a term linear in frequency. When we perform the frequency integral over the response function weighted by this factor, we find that the integrals vanish since one can close in the half-plane where the response function is analytic. (This is the upper half-plane for $G_{\rho B}$ and the lower half-plane for $G_{B\rho}$.) Thus,
\be
G_{B}^{(2,2)}(1)=0.
\ee

Returning to the $\rho$-component of Eq.(\ref{eq:25}), we have
\be
G_{\rho}^{(2,2)}(1)
=-\frac{1}{2}G_{\rho x}^{(0)}(1\bar{2})
\gamma_{xyz}^{(0)}(\bar{2}\bar{3}\bar{4})
\tilde{G}_{yz}(\bar{3}\bar{4}).
\ee
The integrals over the response components of $\tilde{G}$ vanish for the same reasons given above and one is left with the contribution
\be
G_{\rho}^{(2,2)}(1)&=&
-\frac{1}{2}G_{\rho B}^{(0)}(10)
\int d3 \gamma_{B\rho\rho}^{(0)}(0,3,-3)
\tilde{G}_{\rho\rho}(3)
\nonumber\\
&=&-\frac{1}{2}(-\beta G_{\rho}^{(0)}(1))
\int d3 \gamma_{B\rho\rho}^{(0)}(0,3,-3)
\tilde{G}_{\rho\rho}(3).
\ee
Then, using
\be
\gamma_{B\rho\rho}^{(0)}(0,3,-3)=1/\beta\rho_0^2,
\ee
we are left with
\be
G_{\rho}^{(2,2)}(1)
&=&\delta (1)\frac{1}{2\rho_0}\int d3 \tilde{G}_{\rho\rho}(3)
\nonumber\\
&=&\delta (1)\frac{1}{2\rho_0}\int \frac{d^dk}{(2\pi)^d}\frac{d\omega}{2\pi} \tilde{G}_{\rho\rho}(k,\omega)
\nonumber\\
&=& \delta (1)\rho_0\beta^2\frac{1}{2}\int \frac{d^dk}{(2\pi)^d} V^2(k)S(k)
\label{eq:33a}
\ee
where we use the result
\be
\int \frac{d\omega}{2\pi} \tilde{G}_{\rho\rho}(k,\omega)=
\rho_0^2\beta^2 V^2(k)S(k)
\ee
derived in Appendix \ref{app:staticG}.

Collecting all the contributions to the equation of state, -- Eqs. (\ref{eq:13}), (\ref{eq:18}), (\ref{eq:23}), and (\ref{eq:33a}), -- we have
\be
\bar{\rho}&=&\rho_{0}\left(1-\bar{\rho}\beta V(0)
+\frac{1}{2}(\bar{\rho}\beta V(0))^{2}
+\frac{\beta^2}{2}\int \frac{d^dk}{(2\pi)^d} V^2(k)S(k)
\right)
\nonumber\\
&=&\rho_{0}\left(1-\tilde{V}(0)
+\frac{1}{2}\tilde{V}^2(0)
+\frac{1}{2\bar{\rho}}\int \frac{d^dk}{(2\pi)^d} \tilde{V}^2(k)\tilde{S}(k)
\right)
\label{eq:37}
\ee
where $\tilde{V}=\bar{\rho}\beta V$ and $\tilde{S}=S/\bar{\rho}$.
This agrees with a strictly static formulation of the problem\cite{Leibowitz}, and can be rewritten in the form
\be
\frac{\bar{\rho}}{\rho_0} &=& \exp \bigg[-\tilde{V}(0)
+\frac{1}{2\bar{\rho}}\int \frac{d^d k}{(2\pi)^d} \tilde{V}^2(k)\tilde{S}(k)\bigg]
\ee
Expanding in $\tilde{V}$ leads back to Eq.(\ref{eq:37}).

We elaborate on the equation of state and connect it to a more conventional form in Appendix \ref{app:EOS}.

\subsection{Treatment of ${\cal G}_{ij}$}
We next need to treat the propagator
\be
\mathcal{G}_{ij} = \textrm{Tr}~\phi_i\phi_je^{H\cdot\phi+\Delta W[H]}
\ee
in order to determine $\gamma_{ij}$ using Eq.(\ref{eq:4}). We shall see that $\mathcal{G}_{ij}$ is, roughly speaking, a single-particle quantity. The expansion in powers of $V$ follows closely the expansion in treating the equations of state. After expanding $\Delta W$ in powers of $V$, we have
\be
{\cal G}_{ij}=G_{ij}^{(0)}+G_{ijk}^{(0)}\sigma_{k\ell}G_{\ell}
+\frac{1}{2}G_{ijk\ell}^{(0)}\sigma_{ku}\sigma_{\ell v}(G_{uv}+G_{u}G_{v})
+\mathcal{O}(V^3).
\ee

The first term is just the noninteracting matrix propagator determined in FTSPD as
\be
G_{ij}^{(0)}(12) = G_{ij}^{(0)}(1)\delta(1+2)
\ee
with
\be
G_{\rho B}^{(0)}(1) = \bigg(G_{B\rho}^{(0)}(1)\bigg)^*= \frac{-\beta\rho_0\kappa_1}{\kappa_1-i\omega_1}
\ee
and
\be
G_{\rho\rho}^{(0)}(1) = \frac{2\rho_0\kappa_1}{\omega^2_1+\kappa_1^2}.
\ee
The first- and second-order terms are addressed in turn.

\subsubsection{First-order}
For the first-order contribution,
\be
{\cal G}^{(1)}_{ij}=G_{ijk}^{(0)}\sigma_{k\ell}G_{\ell},
\ee
we again replace the zeroth-order three-point cumulant with the three-point vertex using Eq.(\ref{eq:22}) and find
\be
{\cal G}^{(1)}_{ij}&=&G_{iu}^{(0)}(-\Sigma_{uv}^{(1)})G_{vj}^{(0)}
\label{eq:53}
\ee
where the self-energy is given in Fourier-space by,
\be
\Sigma_{uv}^{(1)}(12)
&=&\gamma_{uvw}^{(0)}(12\bar{3})G_{wk}^{(0)}(\bar{3}\bar{4})\sigma_{k\ell}(\bar{4}\bar{5})G_{\ell}(\bar{5})
\nonumber\\
&=&\gamma_{uvw}^{(0)}(12\bar{3})G_{wB}^{(0)}(\bar{3}0)G_{\rho}(0)V(0)
\nonumber\\
&=&\gamma_{uvw}^{(0)}(12\bar{3})(-\beta G_w^{(0)}(\bar{3})\bar{\rho})V(0)
\nonumber\\
&=&-\beta V(0)\bar{\rho}(-\gamma_{uv}^{(0)}(12))
\nonumber\\
&=&\beta \bar{\rho}V(0)\gamma_{uv}^{(0)}(12).
\ee
Putting this back into Eq.(\ref{eq:53}) and remembering that $G_{iu}^{(0)}\gamma_{uv}^{(0)}G_{vj}^{(0)}=G_{ij}^{(0)}$, we have
\be
\mathcal{G}_{ij}^{(1)}(q,\omega)=-G_{ij}^{(0)}(q,\omega)\beta V(0)\bar{\rho}.
\ee

Combined with the zeroth-order result, we have
\be
\mathcal{G}_{ij}^{(0)}(q,\omega)+\mathcal{G}_{ij}^{(1)}(q,\omega)=G_{ij}^{(0)}(q,\omega)[1-V(0)\beta\bar{\rho}]
\ee
and we see that the first-order contribution is a static contribution to the equation of state.

\subsubsection{Second-order}
Working at second-order we have the two-pieces contributing to $\mathcal{G}_{ij}$,
\be
{\cal G}_{ij}^{(2)}={\cal G}_{ij}^{(2,1)}
+{\cal G}_{ij}^{(2,2)}
\ee
where
\be
{\cal G}_{ij}^{(2,1)}=
\frac{1}{2}G_{ijk\ell}^{(0)}\sigma_{ku}\sigma_{\ell v}G_{u}G_{v}
\ee
and
\be
{\cal G}_{ij}^{(2,2)}
=\frac{1}{2}G_{ijk\ell}^{(0)}\sigma_{ku}\sigma_{\ell v}G_{uv}.
\ee

Let us take the disconnected piece first.  Using the representation
$G_{u}(1)=\bar{\rho}\delta (1)\delta_{u\rho}$,
we have
\be
{\cal G}_{ij}^{(2,1)}(12)=
\frac{1}{2}G_{ijBB}^{(0)}(12 0 0)(V(0)\bar{\rho})^{2}
\nonumber
\ee
\be
=\frac{1}{2}\beta^2 G_{ij}^{(0)}(12)(V(0)\bar{\rho})^{2}
\ee
where we use the identity from Appendix \ref{app:ID} given by
\be
G_{ijBB}^{(0)}(12 0 0)
=\beta^2G_{ij}^{(0)}(12).
\ee

Turning to the substantial contribution, we have
\be
{\cal G}_{ij}^{(2,2)}
=\frac{1}{2}G_{ijk\ell}^{(0)}\sigma_{ku}\sigma_{\ell v}G_{uv}.
\label{eq:63}
\ee
For the first time we encounter the noninteracting four-point cumulant,
\be
G_{ijk\ell}^{(0)}=G_{ix}^{(0)}G_{jy}^{(0)}
G_{kz}^{(0)}G_{\ell w}^{(0)}\gamma_{xyzw}^{(0)},
\label{eq:64}
\ee
where we have introduced the four-point vertex $\gamma_{xyzw}^{(0)}$. It is important to recognize that the four-point vertex has a one-particle reducible contribution and a one-particle irreducible contribution,
\be
\gamma_{xyzw}^{(0)}=\gamma_{xyzw}^{(0,R)}+\gamma_{xyzw}^{(0,I)}.
\label{eq:65}
\ee
The reducible contribution can be written quite generally as
\be
\gamma_{xyzw}^{(0,R)}=\gamma_{xyk}^{(0)}\mathcal{G}_{k\ell}\gamma_{\ell zw}^{(0)}
+\gamma_{xzk}^{(0)}\mathcal{G}_{k\ell}\gamma_{\ell yw}^{(0)}
+\gamma_{xwk}^{(0)}\mathcal{G}_{k\ell}\gamma_{\ell yz}^{(0)}
\ee
where, to the order we are considering here, we may take $\mathcal{G}_{ij}\rightarrow G^{(0)}_{ij}$ such that
\be
\gamma_{xyzw}^{(0,R)}=\gamma_{xyk}^{(0)}G_{k\ell}^{(0)}\gamma_{\ell zw}^{(0)}
+\gamma_{xzk}^{(0)}G_{k\ell}^{(0)}\gamma_{\ell yw}^{(0)}
+\gamma_{xwk}^{(0)}G_{k\ell}^{(0)}\gamma_{\ell yz}^{(0)}.
\label{eq:66}
\ee
The non-interacting four-point irreducible vertex functions are determined in Ref. \onlinecite{cumvert} and summarized in Appendix \ref{app:Vertices}.

Using Eqs.(\ref{eq:65}) and (\ref{eq:66}) in Eq.(\ref{eq:64}), we obtain the second-order contributions to ${\cal G}_{ij}$,
\be
{\cal G}^{(2,2)}_{ij}={\cal G}^{(2,2,R,1)}_{ij}
+{\cal G}^{(2,2,R,2)}_{ij}
+{\cal G}^{(2,2,R,3)}_{ij}
+{\cal G}^{(2,2,I)}_{ij}
\label{eq:56}
\ee
where we have four pieces. The first is a disconnected contribution
\be
{\cal G}^{(2,2,R,1)}_{ij}=
\frac{1}{2}G_{ix}^{(0)}
G_{jy}^{(0)}\gamma_{xy\ell}^{(0)}G_{\ell s}^{(0)}
\gamma_{szw}^{(0)}\tilde{G}_{zw}
\ee
where $\tilde{G}$ is defined by Eq.(\ref{eq:29}). The next two terms are one-loop contributions given by
\be
{\cal G}^{(2,2,R,2)}_{ij}=
\frac{1}{2}G_{ix}^{(0)}
\gamma_{xz\ell}^{(0)}
\tilde{G}_{zw}
G_{\ell s}^{(0)}
\gamma_{wsy}^{(0)}
G_{yj}^{(0)}
\ee
and ${\cal G}^{(2,2,R,3)}_{ij}={\cal G}^{(2,2,R,2)}_{ij}.$ Finally, we have a Hartree-like contribution
\be
{\cal G}^{(2,2,I)}_{ij}=
\frac{1}{2}G_{ix}^{(0)}
G_{jy}^{(0)}\gamma_{xyzw}^{(0,I)}
\tilde{G}_{zw}.
\ee

Consider first the disconnected term
\be
{\cal G}^{(2,2,R,1)}_{ij}(12)=
\frac{1}{2}G_{ix}^{(0)}(1\bar{3})
G_{jy}^{(0)}(2\bar{4})\gamma_{xy\ell}^{(0)}(\bar{3}\bar{4}\bar{5})
G_{\ell s}^{(0)}(\bar{5}\bar{6})
\gamma_{szw}^{(0)}(\bar{6}\bar{7}\bar{8})
\tilde{G}_{zw}(\bar{7}\bar{8}).
\ee
Since the frequency integral over the response components
of $\tilde{G}_{zw}$ vanish we have
\be
{\cal G}^{(2,2,R,1)}_{ij}(12)
&=&\frac{1}{2}G_{ix}^{(0)}(1\bar{3})
G_{jy}^{(0)}(2\bar{4})\gamma_{xy\rho}^{(0)}(\bar{3}\bar{4}\bar{5})
G_{\ell s}^{(0)}(\bar{5}\bar{6})
\gamma_{s\rho \rho}^{(0)}(\bar{6}\bar{7}\bar{8})
\tilde{G}_{\rho \rho}(\bar{7}\bar{8})
\nonumber\\
&=&\frac{1}{2}G_{ix}^{(0)}(1\bar{3})
G_{jy}^{(0)}(2\bar{4})\gamma_{xy\rho}^{(0)}(\bar{3}\bar{4}\bar{5})
G_{\ell B}^{(0)}(\bar{5}\bar{6})
\gamma_{B\rho \rho}^{(0)}(\bar{6}\bar{7}\bar{8})
\tilde{G}_{\rho \rho}(\bar{7}\bar{8})
\nonumber\\
&=&\frac{1}{2}G_{ix}^{(0)}(1\bar{3})
G_{jy}^{(0)}(2\bar{4})\gamma_{xy\rho}^{(0)}(\bar{3}\bar{4}\bar{5})
G_{\rho B}^{(0)}(\bar{5}\bar{6})
\gamma_{B\rho \rho}^{(0)}(\bar{6}\bar{7}\bar{8})
\tilde{G}_{\rho \rho}(\bar{7}\bar{8}).
\nonumber
\ee

Let us pause and look at the combination
$G_{\rho B}^{(0)}(5\bar{6})\gamma_{B\rho\rho}^{(0)}(\bar{6}\bar{7}\bar{8})\tilde{G}_{\rho\rho}(\bar{7}\bar{8})$. Using the implicit delta functions, we can write this as
\be
G_{\rho B}^{(0)}(5\bar{6})\gamma_{B\rho\rho}^{(0)}(\bar{6}\bar{7}\bar{8})\tilde{G}_{\rho\rho}(\bar{7}\bar{8})
&=& G_{\rho B}^{(0)}(5, 0)\int d7 \gamma_{B\rho\rho}^{(0)}(0,7,-7)\tilde{G}_{\rho\rho}(7)
\ee
Because $\gamma_{B\rho \rho}^{(0)}(0,7,-7)=1/\beta\rho_0^2$, the only remaining integration is over $\tilde{G}_{\rho\rho}(7)$. Inserting this, along with $G_{\rho B}^{(0)}(5,0)=-\beta G_{\rho}^{(0)}(5)$, we have
\be
G_{\rho B}^{(0)}(5\bar{6})\gamma_{B\rho\rho}^{(0)}(\bar{6}\bar{7}\bar{8})\tilde{G}_{\rho\rho}(\bar{7}\bar{8})
&=& (-\beta G_{\rho}^{(0)}(5))\int d7 \tilde{G}_{\rho\rho}(7)/\beta\rho_0^2
\ee
which gives
\be
{\cal G}^{(2,2,R,1)}_{ij}(12)&=&
\frac{1}{2}G_{ix}^{(0)}(1\bar{3})
G_{jy}^{(0)}(2\bar{4})\gamma_{xy\rho}^{(0)}(\bar{3}\bar{4}\bar{5})
(- G_{\rho}^{(0)}(\bar{5}))\int d7 \tilde{G}_{\rho\rho}(7)/\rho_0^2.
\ee
If we finally use $\gamma_{xy\rho}^{(0)}(34\bar{5})G_{\rho}^{(0)}(\bar{5})=-\gamma_{xy}^{(0)}(34)$, then we are left with
\be
{\cal G}^{(2,2,R,1)}_{ij}(12)&=&
\frac{1}{2} G_{ix}^{(0)}(1\bar{3})
G_{jy}^{(0)}(2\bar{4})\gamma_{xy}^{(0)}(\bar{3}\bar{4})
\int d7 \tilde{G}_{\rho\rho}(7)/\rho_0^2
\nonumber\\
&=&\frac{1}{2} G_{ij}^{(0)}(12)
\int d7 \tilde{G}_{\rho\rho}(7)/\rho_0^2.
\ee

This term goes into the statics and the determination of the equation of state to second-order. Combining this with the zeroth- and first-order contributions, we have the very simple result
\be
G_{ij}^{(0)}+\mathcal{G}_{ij}^{(1)}+\mathcal{G}_{ij}^{(2,1)}+\mathcal{G}_{ij}^{(2,2,R,1)} &=&
G_{ij}^{(0)}\bigg[1-\bar{\rho}\beta V(0)+\frac{1}{2}(\bar{\rho}\beta V(0))^2
+\frac{\beta^2}{2}\int \frac{d^dk}{(2\pi)^d} V^2(k) S(k)\bigg]
\nonumber\\
&=&G_{ij}^{(0)}\frac{\bar{\rho}}{\rho_0}.
\ee
The remaining three terms in Eq.(\ref{eq:56}) are of the form of self-energy terms which make contributions to $\gamma_{ij}^{(2)}$.

\subsubsection{Summary of results at second-order}

The results for the single-particle propagator to second-order in perturbation theory can be written in the form
\be
{\cal G}_{ij}=
\frac{\bar{\rho}}{\rho_{0}}G_{ij}^{(0)}-G_{ik}^{(0)}\gamma_{k\ell}^{(2)}G_{\ell j}^{(0)}.
\label{eq:55}
\ee
where
\be
\gamma_{ij}^{(2)}&=&
\gamma_{ij}^{(2,loop)}+
\gamma_{ij}^{(2,H)}
\ee
and where
\be
\gamma_{ij}^{(2,loop)}&=&
-\gamma_{ik\ell}^{(0)}
\tilde{G}_{kx}
G_{\ell y}^{(0)}
\gamma_{xyj}^{(0)},
\label{eq:67}
\ee
and
\be
\gamma_{ij}^{(2,H)}
=-\frac{1}{2}
\gamma_{ijk\ell}^{(0,I)}
\tilde{G}_{k\ell}.
\ee

To this order in perturbation theory we can rewrite Eq.(\ref{eq:55}) in the form
\be
\mathcal{G}_{ij} = \bar{G}_{ij}^{(0)} - \bar{G}_{ik}^{(0)}\bar{\gamma}_{k\ell}^{(2)}\mathcal{G}_{\ell j}
\ee
where $\bar{G}^{(0)}_{ij}$ is $G_{ij}^{(0)}$ with $\rho_0$ replaced by $\bar{\rho}$ and $\bar{\gamma}^{(2)}_{ij}$ is $\gamma_{ij}^{(2)}$ with $\rho_0$ replaced by $\bar{\rho}$. Comparing with Eq.(\ref{eq:4}), we can identify
\be
\gamma_{ij} = \bar{\gamma}_{ij}^{(0)}+\bar{\gamma}_{ij}^{(2)}.
\ee
There is no explicit first-order term for the single-particle two-point vertex.

\subsection{The full two-point vertex function}
The two-point vertex is then the sum of the single-particle contribution $\gamma_{ij}$ and the collective contribution $K_{ij}$,
\be
\Gamma_{ij} = \gamma_{ij}+K_{ij}.
\ee
Let us review the collective contributions and discuss the full vertex function.

The first order collective contribution is simply
\be
K_{ij}^{(1)} = -\sigma_{ij},
\ee
while the details of the second order collective contribution are worked out carefully in SDENE and result in
\be
K_{ij}^{(2)} = -\frac{1}{2}\gamma_{iuv}\bar{G}_{uw}\bar{G}_{vz}\Gamma_{wzj}
\ee
where
\be
\bar{G}_{\rho\rho}=\mathcal{G}_{\rho x}\sigma_{xy}G_{y\rho}.
\ee
In SDENE, we kept terms of the lowest order in the vertices and effective cumulants to develop the nontrivial approximation
\be
K_{ij}^{(2)} = -\frac{1}{2}\gamma^{(0)}_{iuv}\bar{G}_{uw}\bar{G}_{vz}\gamma^{(0)}_{wzj}
\ee
with the symmetrized propagator
\be
\bar{G}_{ij} = \frac{1}{2}(G_{ik}^{(0)} \sigma_{k\ell} G_{\ell j} + G_{ik}\sigma_{k\ell} G_{\ell j}^{(0)}).
\ee

Bringing all these pieces together, we have for the second-order two-point vertex,
\be
\Gamma_{ij}(1) = -\bar{\gamma}_{ij}^{(0)}(1)+K_{ij}^{(1)}(1)
+\bar{\gamma}_{ij}^{(2)}(1) +K_{ij}^{(2)}(1).
\ee

Notice that $\bar{\gamma}_{ij}^{(2,loop)}$ and $K_{ij}^{(2)}$ share the same one-loop structure, but with different propagators. All the propagators -- $G_{ij}$, $G^{(0)}_{ij}$, $\bar{G}_{ij}$, and $\tilde{G}_{ij}$ -- satisfy the fluctuation-dissipation relation.

It was shown in SDENE that loop contributions like these can be written as
\be
\gamma_{ij}^{(2)} = 2\hat{\mathcal{O}}^{(\tilde{G}G^{(0)})}[J_{ij}]
\label{eq:79}
\ee
and
\be
K_{ij}^{(2)} = \hat{\mathcal{O}}^{(\bar{G}\bar{G})}[J_{ij}]
\label{eq:80}
\ee
where
\be
\hat{\mathcal{O}}^{(\bar{G}\bar{G})}[J_{ij}]
&=&\frac{1}{\bar{\rho}^4}\int\frac{d^dk_3}{(2\pi)^d}\frac{d^dk_4}{(2\pi)^d}
(2\pi)^d\delta(q_1-k_3-k_4)
\nonumber\\
&&\times\int\frac{d\omega_3}{2\pi}\frac{d\omega_4}{2\pi}
\bar{G}(q_3,\omega_3)\bar{G}(q_4,\omega_4)J_{ij},\\
\hat{\mathcal{O}}^{(\tilde{G}G^{(0)})}[J_{ij}]
&=&\frac{1}{\bar{\rho}^4}\int\frac{d^dk_3}{(2\pi)^d}\frac{d^dk_4}{(2\pi)^d}
(2\pi)^d\delta(q_1-k_3-k_4)
\nonumber\\
&&\times\int\frac{d\omega_3}{2\pi}\frac{d\omega_4}{2\pi}
\tilde{G}(q_3,\omega_3)G^{(0)}(q_4,\omega_4)J_{ij},\\
J_{B\rho} &=& \frac{1}{2\beta} + \frac{\omega_1}{2\beta}\frac{(1+i(\omega_3\bar{K}_{13}+\omega_4\bar{K}_{14}))^2}{\omega_3+\omega_4-\omega_1-i\eta},
\ee
and
\be
J_{BB} = -\textrm{Im}\frac{1}{\beta^2}\frac{(1+i(\omega_3\bar{K}_{13}+\omega_4\bar{K}_{14}))^2}{\omega_3+\omega_4-\omega_1-i\eta}.
\ee

Let us look at the static limit where the structure factor is related to the potential in perturbation theory by
\be
S(q) = -\frac{1}{\beta\Gamma_{B\rho}(q,0)}.
\ee

The terms up to first order are easy to simplify in the $\omega\rightarrow 0$ limit and we have
\be
\bar{\gamma}_{B\rho}^{(0)}(q,0) = -\frac{1}{\beta\bar{\rho}}
\ee
and
\be
K_{B\rho}^{(1)}(q,0) = -V(q).
\ee

Looking next at the second order terms, we have for the collective loop term
\be
K_{B\rho}^{(2)}(q,0)&=&\frac{1}{2\beta\bar{\rho}^4}\int\frac{d^dk_3}{(2\pi)^d}\frac{d\omega_3}{2\pi}
\bar{G}_{\rho\rho}(q_3,\omega_3)\int\frac{d\omega_4}{2\pi}\bar{G}_{\rho\rho}(q-k_3,\omega_4)
\nonumber\\
&=&\frac{1}{2\beta\bar{\rho}^4}\int\frac{d^dk_3}{(2\pi)^d}(-\bar{\rho}\beta V(k_3)S(k_3))(-\bar{\rho}\beta V(q-k_3)S(q-k_3))
\nonumber\\
&=&\frac{1}{2\beta\bar{\rho}^2}\int\frac{d^dk}{(2\pi)^d}\tilde{V}(k)\tilde{S}(k)\tilde{V}(q-k)\tilde{S}(q-k)
\ee
where we have used the zero time results derived in SDENE.

Doing the same with the single-particle loop contribution, we find
\be
\gamma_{B\rho}^{(2,loop)}(q,0) &=& \frac{1}{\bar{\rho}^4}
\int\frac{d^dk_3}{(2\pi)^d}
\int\frac{d\omega_3}{2\pi}\tilde{G}_{\rho\rho}(k_3,\omega_3)
\int\frac{d\omega_4}{2\pi}\bar{G}_{\rho\rho}^{(0)}(q-k_3,\omega_4)
\nonumber\\
&=&\frac{1}{\bar{\rho}^3}\int\frac{d^dk_3}{(2\pi)^d}\int\frac{d\omega_3}{2\pi}
\tilde{G}_{\rho\rho}(k_3,\omega_3)
\nonumber\\
&=&\frac{1}{\beta\bar{\rho}^2}
\int\frac{d^dk_3}{(2\pi)^d}\tilde{V}^2(q)\tilde{S}(q)
\ee

We now look at the Hartree-like term. Using the results summarized in Appendix \ref{app:Vertices}, we have
\be
\gamma_{B\rho}^{(2,H)}(q,\omega)
&=&-\frac{1}{2}\bar{\gamma}_{B\rho k\ell}^{(0,I)}\tilde{G}_{k\ell}
\nonumber\\
&=&-\frac{1}{2}\bigg[\gamma_{B\rho\rho\rho}^{(0,I)}\tilde{G}_{\rho\rho}
+\gamma_{B\rho B\rho}^{(0,I)}\tilde{G}_{B\rho}+\gamma_{B\rho\rho B}^{(0,I)}\tilde{G}_{\rho B}\bigg]
\nonumber\\
&=&-\frac{1}{2\beta\bar{\rho}^3}\int\frac{d^dk_3}{(2\pi)^d}\frac{d\omega_3}{2\pi}
\bigg[\bigg(2+\frac{i\omega\alpha_{2}}{\kappa_1}-\frac{i\omega_3\alpha_3}{\kappa_3}
+\frac{i\omega_3\alpha_{4}}{\kappa_3}\bigg)\tilde{G}_{\rho\rho}(k_3,\omega_3)
\nonumber\\
&&-\frac{2}{\beta}\frac{\alpha_3}{\kappa_3}\tilde{G}_{B\rho}(k_3,\omega_3)
-\frac{2}{\beta}\frac{\alpha_4}{\kappa_3}\tilde{G}_{\rho B}(k_3,\omega_3)\bigg]
\ee
where (under our constraints)
\be
\alpha_{2}&=&\bar{K}_{1+3,1}K_{1+3,-1}+\bar{K}_{1-3,1}K_{1-3,-1},\\
\alpha_{3}&=&\bar{K}_{1-3,1}K_{1-3,3},\\
\alpha_{4}&=&\bar{K}_{1+3,1}K_{1+3,-3}.
\ee
Noting that integrals over $\omega_3\tilde{G}_{\rho\rho}(\omega_3)$ vanish due to odd symmetry and integrals over the response functions vanish when the contour is closed in the appropriate half plane, we are left with
\be
\gamma_{B\rho}^{(2,H)}(q,\omega)
&=&-\frac{1}{\beta\bar{\rho}^2}
\int\frac{d^dk_3}{(2\pi)^d}\tilde{V}^2(k_3)\tilde{S}(k_3)
\nonumber\\
&&+\frac{i\omega}{2\beta\bar{\rho}^2}
\int\frac{d^dk_3}{(2\pi)^d}\bigg(\frac{\alpha_2}{\kappa_1}\bigg)\tilde{V}^2(k_3)\tilde{S}(k_3)
\ee

We can take a closer look at the second term. Writing out the full form of $\alpha_2$, we have
\be
&&\frac{i\omega}{2\beta\bar{\rho}^2}
\int\frac{d^dk_3}{(2\pi)^d}\bigg(\frac{\alpha_2}{\kappa_1}\bigg)\tilde{V}^2(k_3)\tilde{S}(k_3)
\nonumber\\&=&\frac{-i\omega}{2\beta\bar{\rho}^2}
\int\frac{d^dk_3}{(2\pi)^d}\bigg(\frac{(q^2+q\cdot k_3)^2}{(q+k_3)^2q^2}+\frac{(q^2-q\cdot k_3)^2}{(q-k_3)^2q^2}\bigg)
\tilde{V}^2(k_3)\tilde{S}(k_3)
\ee
If we perform a change of variables,
\be
p^2 &=& (q-k)^2 = q^2+k^2-2q\cdot k\\
v^2 &=& (q+k)^2 = q^2+k^2+2q\cdot k,
\ee
then the measures of the two terms change to
\be
d^3k_3 = dk_3 d\phi du k_3^2 = -\frac{2\pi k_3p}{q} dk_3 dp = \frac{2\pi k_3v}{q} dk_3 dv
\ee
where $u = \cos(\theta)$ and $du = -\sin(\theta)d\theta$ is the standard angular integration substitution. Using this set of variables, it is easy to see that the term vanishes,
\be
\int\frac{d^dk_3}{(2\pi)^d}\bigg(\frac{\alpha_2}{\kappa_1}\bigg)\tilde{V}^2(k_3)\tilde{S}(k_3)
&=&\int\frac{dk_3}{2\pi}\tilde{V}^2(k)\tilde{S}(k)
\bigg(-\int\frac{dv}{2\pi}\frac{k_3v}{q}\frac{(v^2+q^2-k^2)^2}{4v^2q^2}
\nonumber\\
&&+\int\frac{dp}{2\pi}\frac{k_3p}{q}\frac{(p^2+q^2-k^2)^2}{4p^2q^2}\bigg) = 0.
\ee

This leaves us with an $\omega$-independent Hartree term given by
\be
\gamma_{B\rho}^{(2,H)}(q,\omega)
&=&-\frac{1}{\beta\bar{\rho}^2}\int\frac{d^dk_3}{(2\pi)^d}\tilde{V}^2(k_3)\tilde{S}(k_3)
\label{eq:91}
\ee
which cancels the loop term:
\be
\gamma_{B\rho}^{(2)}(q,0) = 0.
\ee

Collecting the results, we have finally
\be
\Gamma_{B\rho}(q,0)=-\frac{1}{\beta\bar{\rho}}-V(q)
+\frac{1}{2\beta\bar{\rho}^2}\int\frac{d^dk}{(2\pi)^d}\tilde{V}(k)\tilde{S}(k)\tilde{V}(q-k)\tilde{S}(q-k).
\ee
or, inverting,
\be
\tilde{S}^{-1}(q) = 1+\tilde{V}(q)-
\frac{1}{2\bar{\rho}}\int\frac{d^dk}{(2\pi)^d}\tilde{V}(k)\tilde{S}(k)\tilde{V}(q-k)\tilde{S}(q-k).
\label{eq:94}
\ee
This is the same quantity evaluated in SDENE to determine the effective potential.

\section{The kinetic equation}
\subsection{The kinetic equation}
Having determined $\Gamma_{ij}$ to second-order, we could proceed to solve the Dyson's equation for $G_{\alpha\beta}$. However, there is a more economical route sketched out in SDENE that takes advantage of the FDR and which is nonperturbative. Using the FDR, we can go from Dyson's equations to a single equation for $G_{\rho\rho}(k,t)$. We fill in the details of the analysis given in SDENE.

To derive the kinetic equation, we begin with the $B\rho$ component of Dyson's equation:
\be
\Gamma_{Bx}G_{x\rho} = \Gamma_{BB}G_{B\rho} + \Gamma_{B\rho}G_{\rho\rho} = \delta_{B\rho} = 0.
\ee
In $q$-, $t$-space, this is explicitly
\be
\int ds~\Gamma_{B\rho}(q,t-s)G_{\rho\rho}(q,s-t')
+\int ds~\Gamma_{BB}(q,t-s)G_{B\rho}(s-t')=0.
\label{eq:74}
\ee

Let us split the two-point vertex into two contributions as
\be
\Gamma_{ij}(q,t) = \gamma_{ij}^{(1)}(q,t)-\Sigma_{ij}(q,t),
\ee
where we define $\gamma_{ij}^{(1)}(q,t)$ to be all terms local in time such that
\be
\gamma^{(1)}_{ik}(q,t-s)G_{kj}(q,s-t') = \gamma^{(1)}_{ik}(q,t)G_{kj}(q,t-t').
\ee
From this, it follows that
\be
\gamma^{(1)}_{ij} = \bar{\gamma}_{ij}^{(0)}
+ K^{(1)}_{ij}+\gamma_{ij}^{(2,H)}.
\ee
The second group, $\Sigma_{ij}(q,t)$, is the dynamic memory function\cite{MemFunc} which retains its convolution form and is made up of the remaining contributions,
\be
\Sigma_{ij}(q,t) = -\gamma_{ij}^{(2,loop)}-K_{ij}^{(2)}.
\ee

We may now write Eq.(\ref{eq:74}) as
\be
\gamma^{(1)}_{B\rho}(t)G_{\rho\rho}(t,t')
+\gamma^{(1)}_{BB}(t)G_{B\rho}(t,t') =\Psi (t,t')
\label{eq:107}
\ee
where
\be
\Psi (t,t')=
\int_{-\infty}^{t}ds \Sigma_{B\rho}(t-s)G_{\rho\rho}(s-t')
+\int_{-\infty}^{t'}ds
\Sigma_{BB}(t-s)G_{B\rho}(s-t')
\ee
using the fact that $\Sigma_{B\rho}(t-s)\sim \theta (t-s)$ and $G_{B\rho }(s-t')\sim \theta (t'-s)$.  We then use the
fluctuation-dissipation relations
\be
\Sigma_{B\rho}(t-s)=\theta (t-s)\beta\frac{\partial}{\partial t}
\Sigma_{BB}(t-s)
\ee
and
\be
G_{B\rho}(s-t')=\theta (t'-s)\beta\frac{\partial}{\partial t'}
G_{\rho\rho}(s-t')
\ee
to obtain
\be
\Psi (t,t')=
-\int_{-\infty}^{t}ds \left[\beta\frac{\partial}{\partial s}
\Sigma_{BB}(t-s)\right]
G_{\rho\rho}(s-t')
-\int_{-\infty}^{t'}ds
\Sigma_{BB}(t-s)
\beta\frac{\partial}{\partial s}
G_{\rho\rho}(s-t').
\ee
If we integrate the first integral by parts, we have
\be
\Psi (t,t')&=&- \beta\Sigma_{BB}(0)G_{\rho\rho}(t-t')
+\beta\int_{-\infty}^{t}ds\Sigma_{BB}(t-s)
\frac{\partial}{\partial s}G_{\rho\rho}(s-t')
\nonumber\\
&&-\beta\int_{-\infty}^{t'}ds
\Sigma_{BB}(t-s)
\frac{\partial}{\partial s}
G_{\rho\rho}(s-t')
\nonumber\\
&=&-\beta\Sigma_{BB}(0)G_{\rho\rho}(t-t')
+\beta\int_{t'}^{t}ds \Sigma_{BB}(t-s)
\frac{\partial}{\partial s}
G_{\rho\rho}(s-t')
\ee
where we assume $t > t'$. Putting $\Psi(t-t')$ back into Eq.(\ref{eq:107}) and setting $G_{B\rho}=0$ (due to $t>t'$), we then have the kinetic equation
\be
\gamma^{(1)}_{B\rho}(t)G_{\rho\rho}(t-t')
=-\beta\Sigma_{BB}(0)G_{\rho\rho}(t-t')
+\beta\int_{t'}^{t}ds \Sigma_{BB}(t-s)
\frac{\partial}{\partial s}
G_{\rho\rho}(s-t').
\ee
This is the same form derived in SDENE, but with the division of the vertices more fully defined.

To continue, we need explicit forms for our local and memory function contributions. Collecting terms, we find
\be
\gamma^{(1)}_{B\rho}(q,t)&=&\frac{-1}{\beta\bar{D}\bar{\rho} q^2}\bigg[\frac{\partial}{\partial t}+\bar{D}q^2\bigg]-V(q)
-\frac{1}{\beta\bar{\rho}^2}\int\frac{d^dk}{(2\pi)^d}\tilde{V}^2(k)\tilde{S}(k).
\ee
The $BB$-contribution to the memory function is a bit more complex and we address it next.

\subsection{Memory function}
Let us set up a Fourier transform for the loop pieces which make up the memory function in the form,
\be
\Sigma_{BB}^{(AB)}(q,t)
&=&-\int\frac{d\omega}{2\pi}e^{-i\omega t}\hat{\mathcal{O}}^{(AB)}
\bigg[-\beta^{-2}\textrm{Im}\frac{(1+i(\omega_{3}\bar{K}_{13}+\omega_{4}\bar{K}_{14}))^{2}}
{\omega_{3}+\omega_{4}-\omega-i\eta}\bigg],
\ee
where AB is short for $\bar{G}\bar{G}$ in the case of the collective contribution and $\tilde{G}G^{(0)}$ for the single particle contribution.

We first concentrate on the argument
\be
-\beta^2J_{BB} = \textrm{Im}\frac{(1+i(\omega_{3}\bar{K}_{13}+\omega_{4}\bar{K}_{14}))^{2}}{\omega_{3}+\omega_{4}-\omega-i\eta}.
\ee
If we change variables such that
\be
x = \omega_{3}\bar{K}_{13}+\omega_{4}\bar{K}_{14}
\ee
and
\be
u = \omega - (\omega_3+\omega_4),
\ee
we have
\be
-\beta^2J_{BB} &=& \textrm{Im}\frac{1+2ix-x^2}{-u-i\eta}
\nonumber\\
&=&-\textrm{Im}\frac{-i\eta+i\eta x^2+2ixu+u+2x\eta-ux^2}{u^2+\eta^2}
\nonumber\\
&=&\frac{\eta(1-x^2)}{u^2+\eta^2}-\frac{2xu}{u^2+\eta^2}.
\ee

Returning to our Fourier transform, we have
\be
\Sigma_{BB}^{(AB)}(q,t)
&=&\beta^{-2}\int\frac{d\omega}{2\pi}e^{-i\omega t}\hat{\mathcal{O}}^{(AB)}
\bigg[\frac{\eta(1-x^2)}{u^2+\eta^2}-\frac{2xu}{u^2+\eta^2}\bigg]
\nonumber\\
&=&\beta^{-2}\hat{\mathcal{O}}^{(AB)}
\bigg[e^{-i(\omega_3+\omega_4)t}\int\frac{du}{2\pi}e^{-iu t}
\bigg\{\frac{\eta(1-x^2)}{u^2+\eta^2}-\frac{2xu}{u^2+\eta^2}\bigg\}\bigg]
\nonumber\\
&=&\beta^{-2}\hat{\mathcal{O}}^{(AB)}
\bigg[e^{-i(\omega_3+\omega_4)t}\bigg\{\frac{1}{2}(1-x^2)+ix\bigg\}\bigg].
\ee

At this point, we have now exactly the exponential weight required ($\exp(-i(\omega_3+\omega_4)t)$ to factorize the problem. We have (upon substituting $x$ back in) the explicit result
\be
\Sigma^{(AB)}_{BB}(q_1,t_1)
&=&\frac{1}{2\beta^{2}\bar{\rho}^4}\int \frac{d^d k_3}{(2\pi)^d}\frac{d^d k_4}{(2\pi)^d}
(2\pi)^d\delta (q_{1}-k_{3}-k_{4}) \int\frac{d\omega_3}{2\pi}\frac{d\omega_4}{2\pi}e^{-i(\omega_3+\omega_4)t}
\nonumber\\
&&\times\bigg[1+2i(\omega_3\bar{K}_{13}+\omega_4\bar{K}_{14})
-(\omega_3\bar{K}_{13}+\omega_4\bar{K}_{14})^2\bigg]A(k_3,\omega_3)B(k_4,\omega_4)
\nonumber\\
&=&\frac{1}{2\beta^{2}\bar{\rho}^4}\int \frac{d^d k_3}{(2\pi)^d}\frac{d^d k_4}{(2\pi)^d}
(2\pi)^d\delta (q_{1}-k_{3}-k_{4})\int dt_1^{\prime} \delta (t_1-t_1^{\prime})
\nonumber\\
&&\times\bigg[1-2\bar{K}_{13}\frac{\partial}{\partial t_1}-2\bar{K}_{14}\frac{\partial}{\partial t_1^{\prime}}
+\bar{K}_{13}^2\frac{\partial^2}{\partial t_1^2}+\bar{K}_{14}^2\frac{\partial^2}{\partial t_1^{\prime 2}}
+2\bar{K}_{13}\bar{K}_{14}\frac{\partial}{\partial t_1}\frac{\partial}{\partial t_1^{\prime}}\bigg]
\nonumber\\
&&\times A(k_3,\omega_3)B(k_4,\omega_4)
\ee
where we have introduced an auxiliary variable $t_1^{\prime}$ to help show that the derivatives act only on specific terms.

Moving from generic to our specific loop contributions, we have
\be
\Sigma_{BB}(q_1,t_1)
&=&\frac{1}{2\beta^{2}\bar{\rho}^4}\int \frac{d^d k_3}{(2\pi)^d}\frac{d^d k_4}{(2\pi)^d}
(2\pi)^d\delta (q_{1}-k_{3}-k_{4})\int dt_1^{\prime} \delta (t_1-t_1^{\prime})
\nonumber\\
&&\times\bigg[1-2\bar{K}_{13}\frac{\partial}{\partial t_1}-2\bar{K}_{14}\frac{\partial}{\partial t_1^{\prime}}
+\bar{K}_{13}^2\frac{\partial^2}{\partial t_1^2}+\bar{K}_{14}^2\frac{\partial^2}{\partial t_1^{\prime 2}}
+2\bar{K}_{13}\bar{K}_{14}\frac{\partial}{\partial t_1}\frac{\partial}{\partial t_1^{\prime}}\bigg]
\nonumber\\
&&\times\bigg[2\tilde{G}_{\rho\rho}(k_3,t_1)G^{(0)}_{\rho\rho}(k_4,t_1^{\prime})
+\bar{G}_{\rho\rho}(k_3,t_1)\bar{G}_{\rho\rho}(k_4,t_1^{\prime})\bigg]
\ee
We see that we generate time derivatives because the three-point vertices are frequency dependent.

In addition to the full time-dependent form, we also need the $t=0$ contribution\cite{SigmaBB}. In the limit of small time, the derivative terms vanish and we find the simple result
\be
\Sigma_{BB}(q,t=0)&=&\frac{1}{\beta^2\bar{\rho}^2}
\int\frac{d^dk}{(2\pi)^d}\tilde{V}^2(k)\tilde{S}(k)
\nonumber\\
&&+ \frac{1}{2\beta^2\bar{\rho}^2}\int\frac{d^dk}{(2\pi)^d}\tilde{V}(k)\tilde{S}(k)\tilde{V}(q-k)\tilde{S}(q-k).
\ee
Using our results for the static structure factor to second order given in Eq.(\ref{eq:94}), we can rewrite the second term, giving us
\be
\Sigma_{BB}(q,t=0)&=&\frac{1}{\beta^2\bar{\rho}^2}
\int\frac{d^dk}{(2\pi)^d}\tilde{V}^2(k)\tilde{S}(k)
+ \frac{1}{\beta^2\bar{\rho}}\bigg(1+\tilde{V}(q)-\tilde{S}^{-1}(q)\bigg).
\ee

\subsection{Final form}

Inserting these results into the kinetic equation, we see that a number of terms will cancel yielding
\be
\frac{\partial}{\partial t} G_{\rho\rho}(q,t-t')
&=& -\bar{D} q^2\bar{\rho}S^{-1}(q)G_{\rho\rho}(q,t-t')
\nonumber\\
&&-\bar{D}q^2\int_{t'}^t ds \beta^2\bar{\rho}\Sigma_{BB}(q,t-s)\frac{\partial}{\partial s} G_{\rho\rho}(q,s-t').
\ee
Performing a simple shift in time, we can rewrite this as
\be
\frac{\partial}{\partial t} G_{\rho\rho}(q,t)
&=& -\bar{D} q^2\bar{\rho}S^{-1}(q)G_{\rho\rho}(q,t)
\nonumber\\
&&-\bar{D}q^2\int_{0}^t ds \beta^2\bar{\rho}\Sigma_{BB}(q,t-s)\frac{\partial}{\partial s} G_{\rho\rho}(q,s)
\ee

This equation is of the memory function form where we now have a field-theoretic prescription for the determination of $\Sigma_{BB}(q,t-s)$. The static part of the memory function yields a term proportional to the inverse static structure factor. The dynamic part of the memory function is just the $BB$ matrix element of the loop contributions, $\gamma_{ij}^{(2,loop)}+K_{ij}^{(2)}$. In SDENE we showed explicitly that $K_{ij}^{(2)}$ satisfies a FDR and we will show elsewhere that $\gamma_{ij}^{(2,loop)}$ explicitly satisfies a FDR as well.

\section{Conclusion}
We have shown here that, in the case where one is in thermal equilibrium, the density-density correlation function satisfies a kinetic equation of the same form as in MCT\cite{Crisanti, Goetze, Das}. The interesting point is that we can explore the corrections of the relevant memory function, $\Sigma_{BB}$. Since $\Sigma_{BB}$ comes from a detailed microscopic derivation, there are several features which differ from the conventional mode coupling analysis. At second-order in perturbation theory, we have a structure
\be
\Sigma^{(2)}_{BB}(12) = \frac{1}{2}
\includegraphics{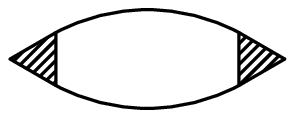}
\ee
where one has a one-loop structure where the three-point vertices have a frequency dependence. There are two pairs of effective propagators. One contribution is a product of $\bar{G}_{ij}$ propagators where
\be
\bar{G}_{ij} = \frac{1}{2}(G_{ik}^{(0)} \sigma_{k\ell} G_{\ell j} + G_{ik}\sigma_{k\ell} G_{\ell j}^{(0)})
\ee
and the other contribution is from the product of $\tilde{G}_{ij}$ and $G_{ij}^{(0)}$ where
\be
\tilde{G}_{ij} = G_{iw}^{(0)}\sigma_{wx} G_{xy}\sigma_{yz} G_{zj}^{(0)}.
\ee

Thus, the microscopic theory is more involved than MCT. In Appendix \ref{app:FDS}, we show that $\bar{G}_{ij}$ and $\tilde{G}_{ij}$ themselves satisfy a FDR.

At the next order in perturbation theory, one generates two-loop structures such as
\be
\Sigma^{(3)}_{BB}(12) & = &
\includegraphics{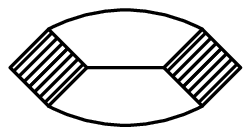}
\ee

In a companion paper\cite{KE_Paul}, we will look at the long-time kinetics generated by the kinetic equation using analytic techniques. The main result is that one finds, as in MCT, that the late time decay shows two power-law regimes governed by exponents $a$ and $b$. We show in the current case that $a$ and $b$ satisfy the relation
\be
\frac{\Gamma(1-a)^2}{\Gamma(1-2a)} = \lambda = \frac{\Gamma(1+b)^2}{\Gamma(1+2b)}
\ee
where $\lambda$ is a parameter determined in the model.

In a second companion paper\cite{KE_Dave}, we look at the numerical evaluation of the second-order kinetic equation derived here.

\begin{acknowledgments}
This work was supported by the Joint Theory Institute and the Department of Physics at the University of Chicago. D. McCowan would also like to acknowledge support from the Department of Education.
\end{acknowledgments}

\appendix
\section{Summary of the zeroth-order three-point vertex functions}
\label{app:Vertices}
Let us begin with a few definitions. First, we have the commonly occurring wavenumber combinations
\be
K_{ij}&=&\bar{D}q_i\cdot q_j,\\
\kappa_i &=& K_{ii} = \bar{D}q_i^2,
\ee
and
\be
\bar{K}_{ij}=\frac{K_{ij}}{\kappa_i\kappa_j}.
\ee
We also have the important combination
\be
G_i^{-1}=-i\omega_i+\kappa_i.
\ee

The two-point vertex functions were first derived in SDENE and are
\be
\gamma^{(0)}_{\rho\rho}(12)&=&0,\\
\gamma^{(0)}_{B\rho}(12)&=&-\frac{G_{1}^{-1}}{\beta\rho_{0}\kappa_{1}}\delta (1+2),\\
\gamma^{(0)}_{\rho B}(12)&=&-\frac{G_{1}^{-1,*}}{\beta\rho_{0}\kappa_{1}}\delta (1+2),
\ee
and
\be
\gamma^{(0)}_{B B}(12)=-\frac{2}{\beta^2\rho_{0}\kappa_{1}}\delta (1+2).
\ee

The three-point vertex functions were first derived in Ref. \onlinecite{cumvert} and are
\be
\gamma_{\rho\rho\rho}^{(0)}(123)&=&0,\\
\gamma_{B\rho\rho}^{(0)}(123)&=&
-\frac{1}{\beta\rho_{0}^{2}}
\left[\bar{K}_{12}G_{2}^{-1,*}
+\bar{K}_{13}G_{3}^{-1,*}\right]
\delta (1+2+3)
\nonumber\\
&=&\frac{1}{\beta\rho_{0}^{2}}[1-iE_{1}]
\delta (1+2+3),\\
\gamma_{BB\rho}^{(0)}(123)&=&
-2\frac{1}{\beta^{2}\rho_{0}^{2}}\bar{K}_{12}
\delta (1+2+3),
\ee
and
\be
\gamma_{BBB}^{(0)}(123)=0
\ee
where
\be
E_1 = \omega_2\bar{K}_{12}+\omega_3\bar{K}_{13}
\ee
and where the other vertices are easily constructed by symmetry.

The four-point vertex functions were also derived in Ref. \onlinecite{cumvert} and are the sum of a reducible and irreducible contribution,
\be
\gamma^{(0)}_{ijk\ell}=\gamma_{ijk\ell}^{(0,R)}+\gamma_{ijk\ell}^{(0,I)},
\ee
where
\be
\gamma^{(0,R)}_{ijk\ell}=\gamma^{(0)}_{ijx}G^{(0)}_{xy}\gamma^{(0)}_{yk\ell}
+\gamma^{(0)}_{ikx}G^{(0)}_{xy}\gamma^{(0)}_{yj\ell}+\gamma^{(0)}_{i\ell x}G^{(0)}_{xy}\gamma^{(0)}_{yjk}.
\ee

The full (amputated) results are
\be
\bar{\gamma}^{(0)}_{\rho\rho\rho\rho}&=&0,\\
\bar{\gamma}^{(0)}_{BBBB}&=&16\beta^{-4}N_T,\\
\bar{\gamma}^{(0)}_{BBB\rho}&=&8\beta^{-3}G_4^{-1*}N_T,\\
\bar{\gamma}^{(0)}_{BB\rho\rho}&=&4\beta^{-2}G_3^{-1*}G_4^{-1*}[N_T+M_{12}],
\ee
and
\be
\bar{\gamma}^{(0)}_{B\rho\rho\rho}=2\beta^{-1}G_2^{-1*}G_3^{-1*}G_4^{-1*}[N_T+M_{12}+M_{13}+M_{14}]
\ee
where
\be
M_T&=&M_{12}+M_{13}+M_{14}+M_{23}+M_{24}+M_{34},\\
M_{12}
&=&\frac{1}{8}\rho_0G_3^{-1}G_4^{-1}K_{12}[G^*_{3+4}(G^*_3+G_4^*)+G_{3+4}(G_3+G_4)]\delta(1+2+3+4),
\ee
and
\be
N_T&=&\frac{1}{4}\rho_0\{K_{12}K_{34}[G^*_{3+4}+G_{3+4}]+K_{13}K_{24}[G^*_{2+4}+G_{2+4}]
\nonumber\\
&&+K_{14}K_{23}[G^*_{2+3}+G_{2+3}]\}\delta(1+2+3+4),
\ee

The irreducible pieces are given by
\be
\gamma^{(0,I)}_{\rho\rho\rho\rho}(1234)&=&0,\\
\gamma^{(0,I)}_{BBBB}(1234)&=&0,\\
\gamma^{(0,I)}_{BBB\rho}(1234)&=&0,\\
\gamma_{BB\rho\rho}^{(0,I)}(1234) &=& -\frac{2}{\beta^2\rho_0^3}
\delta(1+2+3+4)\bigg[\bar{K}_{1,1+4}\bar{K}_{2,1+4}\kappa_{1+4}
+\bar{K}_{1,1+3}\bar{K}_{2,1+3}\kappa_{1+3}\bigg]
\nonumber\\
&=&-\frac{2}{\beta^2\rho_0^3}\delta(1+2+3+4)\bigg[Q_{1+4}\kappa_{1+4}+Q_{1+3}\kappa_{1+3}\bigg]
\ee
and
\be
\gamma_{B\rho\rho\rho}^{(0,I)}(1234)&=&\frac{1}{\beta\rho_0^3}\delta(1+2+3+4)\bigg[2
-i\omega_2(\bar{K}_{1,1+4}\bar{K}_{2,1+4}\kappa_{1+4}+\bar{K}_{1,1+3}\bar{K}_{2,1+3}\kappa_{1+3})
\nonumber\\
&&-i\omega_3(\bar{K}_{1,1+2}\bar{K}_{3,1+2}\kappa_{1+2}+\bar{K}_{1,1+4}\bar{K}_{3,1+4}\kappa_{1+4})
\nonumber\\
&&-i\omega_4(\bar{K}_{1,1+2}\bar{K}_{4,1+2}\kappa_{1+2}+\bar{K}_{1,1+3}\bar{K}_{4,1+3}\kappa_{1+3})\bigg]
\nonumber\\
&=&\frac{1}{\beta\rho_0^3}\bigg[
2-\sum_{i=2}^{4}\alpha_{i}
\frac{i\omega_{i}}{\kappa_{i}}\bigg]
\ee
where
\be
Q_{1+3}&=&\bar{K}_{1,1+3}\bar{K}_{2,1+3},\\
Q_{1+4}&=&\bar{K}_{1,1+4}\bar{K}_{2,1+4},\\
\alpha_{2}&=&\bar{K}_{1+3,1}K_{1+3,2}+\bar{K}_{1+4,1}K_{1+4,2},\\
\alpha_{3}&=&\bar{K}_{1+2,1}K_{1+2,3}+\bar{K}_{1+4,1}K_{1+4,3}
\ee
and
\be
\alpha_{4}&=&\bar{K}_{1+2,1}K_{1+2,4}+\bar{K}_{1+3,1}K_{1+3,4}.
\ee

\section{Reduction identities}
\label{app:ID}

A number of ``reduction identities" are used in this work. These are part of a larger collection which is discussed in Ref. \onlinecite{cumvert}. The proofs of these relations are essentially brute force demonstrations and it will be sufficient to sketch the basics here. The reduction identities fall into two types.

First, zeroth-order cumulants of a particular number of fields can be reduced to a cumulant of a smaller number of fields when the argument of the $B$-field is zero. In this work, we particularly make use of two relations,
\be
G_{iB}^{(0)}(10) = -\beta G^{(0)}_{i}(1)
\ee
and
\be
G_{ijBB}^{(0)}(1200) = \beta^2G^{(0)}_{ij}(12),
\ee
which can be verified by explicitly writing out each cumulant and setting the relevant  variables to zero. (For convergence, we must set the frequencies to zero first, then the wavenumbers.) The calculation is tedious, but straightforward and so is omitted.

Second, zeroth-order vertex functions of a particular number of fields can be reduced to a vertex function of a smaller number of fields when the argument of the $\rho$-field is zero, for example with the common term
\be
\gamma_{xy\rho}^{(0)}(120)=-\gamma_{xy}^{(0)}(12)/\rho_0.
\ee
Again, this is a straightforward exercise.

From these two simple facts, we can derive other useful identities. When one convolves a quantity with a one-point cumulant (either $G_i(1)$ or $G_i^{(0)}(1)$), the implicit constraining delta function will usually cause one or more fields in the product to vanish. For example, a combination which appears several times in our work is
\be
\gamma_{xyz}^{(0)}(12\bar{3})G_z^{(0)}(\bar{3})
&=&\gamma_{xy\rho}^{(0)}(120)G_{\rho}^{(0)}(0)
\nonumber\\
&=&(-\gamma_{xy}^{(0)}(12)/\rho_0)\rho_0
\nonumber\\
&=&-\gamma_{xy}^{(0)}(12).
\ee
As another example, consider the first-order contribution to the equation of state (which appears again as a component of the second-order contribution)
\be
G_{\rho}^{(1)}(1)
&=&
G_{\rho B}^{(0)}(1\bar{2})\sigma_{B\rho}(\bar{2}\bar{3})G_{\rho}(\bar{3})
\nonumber\\
&=&G_{\rho B}^{(0)}(1\bar{2})\sigma_{B\rho}(\bar{2}0)G_{\rho}(0)
\nonumber\\
&=&G_{\rho B}^{(0)}(10)\sigma_{B\rho}(00)G_{\rho}(0)
\nonumber\\
&=&G_{\rho B}^{(0)}(10)V(0)\bar{\rho}
\nonumber\\
&=&-\beta G_{\rho}^{(0)}V(0)\bar{\rho}
\nonumber\\
&=&-\beta\rho_0\bar{\rho}V(0)\delta(1).
\ee

\section{Static contribution from $\tilde{G}_{\rho\rho}(1)$}
\label{app:staticG}
Let us look at the integral
\be
\int \frac{d\omega_1}{2\pi}\tilde{G}_{\rho\rho}(q_1,\omega_1)
&=&\int \frac{d\omega_1}{2\pi}\bigg[G_{\rho B}^{(0)}(1)V(1)G_{\rho B}(1)V(1)G_{\rho\rho}^{(0)}(1)
\nonumber\\
&&+G_{\rho \rho}^{(0)}(1)V(1)G_{B\rho }(1)V(1)G_{B\rho}^{(0)}(1)
\nonumber\\
&&+G_{\rho B}^{(0)}(1)V(1)G_{\rho\rho }(1)V(1)G_{B\rho}^{(0)}(1)\bigg]
\nonumber\\
&=&V^{2}(q_1)\int \frac{d\omega_1}{2\pi}\bigg[
G_{\rho \rho}^{(0)}(1)\bigg(G_{\rho B}^{(0)}(1)G_{\rho B}(1)
+G_{B\rho }(1)G_{B\rho}^{(0)}(1)\bigg)
\nonumber\\
&&+G_{\rho B}^{(0)}(1)G_{\rho \rho}(1)G_{B\rho}^{(0)}(1)\bigg].
\ee
Recall the forms of the zeroth-order cumulants
\be
G_{B\rho}^{(0)}(1)&=&\frac{i\kappa_1\rho_0\beta}{\omega_1-i\kappa_1},\\
G_{\rho B}^{(0)}(1)&=&\frac{-i\kappa_1\rho_0\beta}{\omega_1+i\kappa_1},
\ee
and
\be
G_{\rho\rho}^{(0)}(1)&=&\frac{i\rho_0}{\omega_1+i\kappa_1}
-\frac{i\rho_0}{\omega_1-i\kappa_1}
\ee
where $\kappa_{1} = \bar{D}q_1^2$.

Using these, we have
\be
\int \frac{d\omega_1}{2\pi}\tilde{G}_{\rho\rho}(1)
&=&\rho_0^2V^{2}(q_1)\int \frac{d\omega_1}{2\pi}
\bigg[\beta^2\bigg(\frac{-i\kappa_1}{\omega_1+i\kappa_1}\bigg)
G_{\rho \rho}(1)\bigg(\frac{i\kappa_1}{\omega_1-i\kappa_1}\bigg)
\nonumber\\
&&+\beta\bigg(
\frac{i}{\omega_1+i\kappa_1}
-\frac{i}{\omega_1-i\kappa_1}\bigg)
\nonumber\\
&&\times\left(\frac{-i\kappa_1}{\omega_1+i\kappa_1}
G_{\rho B}(1)
+\frac{i\kappa_1}{\omega_1-i\kappa_1}
G_{B\rho }(1)
\right)\bigg].
\ee

For the $G_{\rho B}$ term, we close the contour integral in the
upper half plane while for the $G_{B\rho}$ term, we close in the lower half plane. This gives
\be
\int \frac{d\omega_1}{2\pi}\tilde{G}_{\rho\rho}(1)
&=&\rho_0^2V^{2}(q_1)\bigg[\frac{2\pi i}{2\pi}(-i)
\frac{-i\kappa_1}{2i\kappa_1}\beta G_{\rho B}(q_1,i\kappa_1)
+\frac{-2\pi i}{2\pi}(i)
\frac{i\kappa_1}{-2i\kappa_1}\beta G_{\rho B}(q_1,-i\kappa_1)
\nonumber\\
&&+\beta^2\int \frac{d\omega_1}{2\pi}
\bigg(\frac{\kappa_1^2}{\omega_1^2+\kappa_1^2}\bigg)G_{\rho \rho}(1)\bigg]
\nonumber\\
&=&\rho_0^2V^{2}(q_1)\bigg[-\frac{1}{2}\beta G_{\rho B}(q_1,i\kappa_1)G_{\rho B}(q_1,-i\kappa_1)
\nonumber\\
&&+\beta^2\int \frac{d\omega_1}{2\pi}\bigg(\frac{\kappa_1^2}{\omega_1^2+\kappa_1^2}\bigg)G_{\rho \rho}(1)\bigg].
\ee

Using the FDR to express $G_{B\rho}$ and $G_{\rho B}$ in terms of $G_{\rho\rho}$, we have
the final result
\be
\int \frac{d\omega_1}{2\pi}\tilde{G}_{\rho\rho}(1)
&=&\rho_0^2\beta^2V^{2}(q_1)\int\frac{d\omega_1}{2\pi}
G_{\rho\rho}(q_1,\omega_1)
\bigg(\frac{-\omega_1}{2}\bigg)\bigg(\frac{1}{i\kappa_1-\omega_1}+
\frac{1}{-i\kappa_1-\omega_1}\bigg)
\nonumber\\
&&+G_{\rho \rho}(q_1,\omega_1)\bigg(\frac{\kappa_1^2}{\omega_1^2+\kappa_1^2}\bigg)\bigg]
\nonumber\\
&=&\rho_0^2\beta^2V^{2}(q_1)\int\frac{d\omega_1}{2\pi}
\frac{\omega_1^{2}+\kappa_1^2}{\omega_1^{2}+\kappa_1^{2}}G_{\rho\rho}(q_1,\omega_1)
\nonumber\\
&=&\rho_0^2\beta^2V^{2}(q_1)\int\frac{d\omega_1}{2\pi} G_{\rho\rho}(q_1,\omega_1).
\ee
The integral over $G_{\rho\rho}$ is simply the static structure factor and we have
\be
\int \frac{d\omega_1}{2\pi}\tilde{G}_{\rho\rho}(1) = \rho_0^2\beta^2V^2(q_1)S(q_1).
\ee

\section{Fluctuation-dissipation relations for $\bar{G}$ and $\tilde{G}$}
\label{app:FDS}
In this appendix we prove that the dressed propagators $\bar{G}$ and $\tilde{G}$ individually satisfy the same fluctuation-dissipation relation that $G$ satisfies. These results hold at all orders of perturbation theory.

\subsection{$\bar{G}$ fluctuation-dissipation symmetry}

Recall the form of $\bar{G}_{ij}$ given by
\be
\bar{G}_{ij}=\frac{1}{2}(G_{ix}^{(0)}\sigma_{xy}G_{yj}+G_{ix}\sigma_{xy}G_{yj}^{(0)})
\ee
Explicitly, this yields
\be
\bar{G}_{\rho B}&=&\frac{1}{2}(G_{\rho B}^{(0)}VG_{\rho B}+G_{\rho B}VG_{\rho B}^{(0)})
= G_{\rho B}^{(0)}VG_{\rho B},\\
\bar{G}_{B\rho}&=&\frac{1}{2}(G_{B\rho}^{(0)}VG_{B\rho}+G_{B\rho}VG_{B\rho}^{(0)})
= G_{B\rho}^{(0)}VG_{B\rho},
\ee
and
\be
\bar{G}_{\rho\rho}=\frac{1}{2}(G_{\rho B}^{(0)}VG_{\rho\rho}+G_{\rho\rho}^{(0)}VG_{B\rho}
+G_{\rho B}VG_{\rho\rho}^{(0)}+G_{\rho\rho}VG_{B\rho}^{(0)}).
\ee

If we write out our contributing terms as real and imaginary components,
\be
G_{\rho B}^{(0)} &=& R_0+iI_0\\
G_{\rho B} &=& R+iI\\
G_{B\rho}^{(0)} &=& R_0-iI_0\\
G_{B\rho} &=& R-iI\\
G_{\rho \rho}^{(0)} &=& -\frac{2}{\omega}I_0\\
G_{\rho \rho} &=& -\frac{2}{\omega}I
\ee
then we have for the imaginary part of $\bar{G}_{\rho B}$,
\be
\textrm{Im} \bar{G}_{\rho B} &=& \frac{1}{2i}(\bar{G}_{\rho B}-\bar{G}_{B\rho})
\nonumber\\
&=&\frac{V}{2i}\bigg[(R_0+iI_0)(R+iI)-(R_0-iI_0)(R-iI)\bigg]
\nonumber\\
&=&V(IR_0+I_0R).
\ee

Looking next at $\bar{G}_{\rho\rho}$, we have
\be
\bar{G}_{\rho\rho}&=&\frac{V}{2}\bigg[(R_0+iI_0)\bigg(-\frac{2}{\omega}I\bigg)+(R-iI_0)\bigg(-\frac{2}{\omega}I_0\bigg)
\nonumber\\
&&+(R+iI)\bigg(-\frac{2}{\omega}I_0\bigg)+(R_0-iI_0)\bigg(-\frac{2}{\omega}I\bigg)\bigg]
\nonumber\\
&=&-\frac{2V}{\omega}(2R_0I+2RI_0).
\ee
This implies, then, the normal fluctuation-dissipation relation,
\be
\bar{G}_{\rho\rho}= -\frac{2}{\omega}\textrm{Im}\bar{G}_{\rho B}.
\ee

\subsection{$\tilde{G}$ fluctuation-dissipation symmetry}

We may repeat the same procedure for $\tilde{G}_{ij}$ given by
\be
\tilde{G}_{ij}=G^{(0)}_{ix}\sigma_{xy}G_{yz}\sigma_{zp}G^{(0)}_{pj}.
\ee
Explicitly, we have
\be
\tilde{G}_{\rho B}&=&G^{(0)}_{\rho B}VG_{\rho B}VG^{(0)}_{\rho B},\\
\tilde{G}_{B \rho}&=&G^{(0)}_{B \rho}VG_{B \rho}VG^{(0)}_{B \rho},
\ee
and
\be
\tilde{G}_{\rho\rho}=G^{(0)}_{\rho\rho}VG_{B\rho}VG^{(0)}_{B\rho}
+G^{(0)}_{\rho B}VG_{\rho\rho}VG^{(0)}_{B\rho}+G^{(0)}_{\rho B}VG_{\rho B}VG^{(0)}_{\rho\rho}.
\ee
Using the same decomposition into real and imaginary components, we have
\be
\textrm{Im} \tilde{G}_{\rho B} &=& \frac{1}{2i}\bigg(\tilde{G}_{\rho B}-\tilde{G}_{B\rho}\bigg)
\nonumber\\
&=& \frac{V^2}{2i}\bigg[(R_0+iI_0)(R+iI)(R_0+iI_0)-(R_0-iI_0)(R-iI)(R_0-iI_0)\bigg]
\nonumber\\
&=& V^2[I(R_0^2-I_0^2)+2RR_0I_0]
\ee
and
\be
\tilde{G}_{\rho\rho}&=&-\frac{2}{\omega}I_0V^2(R_0+iI_0)(R+iI)-V^2(R_0^2+I_0^2)\frac{2}{\omega}I
-\frac{2}{\omega}I_0V^2(R-iI)(R_0-iI_0)
\nonumber\\
&=&-\frac{2}{\omega}V^2\bigg[(R_0^2-I_0^2)I+2I_0R_0R\bigg].
\ee

Therefore, we again get the expected fluctuation-dissipation relation,
\be
\tilde{G}_{\rho\rho}=-\frac{2}{\omega}\textrm{Im}\tilde{G}_{\rho B}.
\ee

\section{Equation of State}
\label{app:EOS}
We have identified the equation of state to second order in the pseudo-potential as
\be
\rho_0 = \rho_0(\bar{\rho}) = \bar{\rho}\exp\bigg[\tilde{V}(0)
-\frac{1}{2\bar{\rho}}\int\frac{d^dk}{2\pi}\tilde{V}^2(k)\tilde{S}(k)\bigg].
\label{eq:E1}
\ee

\subsection{Conventional Form}
To connect this to a more conventional form for the equation of state recall that we are in the grand canonical ensemble and
\be
\ell^d \rho_0 = e^{\beta\mu}
\ee
where $\mu$ is the chemical potential and $\ell$ is some microscopic length. We then have the thermodynamic identity
\be
\frac{\partial P}{\partial \bar{\rho}} = \bar{\rho}\frac{\partial \mu}{\partial \bar{\rho}}
\ee
where $P$ is the pressure.
Starting with
\be
\beta\mu = \ln(\rho_0 \ell^d),
\ee
we have
\be
\frac{\partial (\beta P)}{\partial \bar{\rho}} = \bar{\rho} \frac{\partial (\beta\mu)}{\partial \bar{\rho}}
= \frac{\bar{\rho}}{\rho_0}\frac{\partial \rho_0}{\partial \bar{\rho}}.
\label{eq:A1}
\ee

At first order, everything can be cleanly worked out. Starting with
\be
\rho_0 = \bar{\rho}\exp[\tilde{V}(0)]
\ee
we have
\be
\frac{\partial \rho_0}{\partial \bar{\rho}} &=& \exp[\tilde{V}(0)] + \bar{\rho}\exp[\tilde{V}(0)]
\frac{\partial \tilde{V}(0)}{\partial \bar{\rho}}
\nonumber\\
&=& \exp[\tilde{V}(0)] + \tilde{V}(0)\exp[\tilde{V}(0)]
\nonumber\\
&=& \bigg(1+\tilde{V}(0)\bigg)\exp[\tilde{V}(0)]
\nonumber\\
&=& \frac{\rho_0}{\bar{\rho}}\bigg(1+\tilde{V}(0)\bigg).
\ee
Putting this into Eq.(\ref{eq:A1}), we find
\be
\frac{\partial (\beta P)}{\partial \bar{\rho}} =
\frac{\bar{\rho}}{\rho_0}\frac{\rho_0}{\bar{\rho}}\bigg(1+\tilde{V}(0)\bigg)
=1+\beta V(0)\bar{\rho}.
\ee
Therefore,
\be
\beta P = \bar{\rho} + \frac{1}{2}\beta V(0)\bar{\rho}^2.
\ee
Clearly, we find the ideal gas law and the first order correction.

If we write more generally that
\be
\rho_0 = \bar{\rho}e^{W[\bar{\rho}]}
\label{eq:E10}
\ee
we have
\be
\frac{\partial \rho_0}{\partial \bar{\rho}} &=& \exp[W] + \bar{\rho}\exp[W]\frac{\partial W}{\partial \bar{\rho}}
\nonumber\\
&=&\exp[W] \bigg(1+ \bar{\rho}\frac{\partial W}{\partial \bar{\rho}}\bigg)
\nonumber\\
&=&\frac{\rho_0}{\bar{\rho}} \bigg(1+ \bar{\rho}\frac{\partial W}{\partial \bar{\rho}}\bigg)
\ee
which yields,
\be
\frac{\partial (\beta P)}{\partial \bar{\rho}}
&=&1+ \bar{\rho}\frac{\partial W}{\partial \bar{\rho}}.
\ee

In perturbation theory, we have
\be
W = \tilde{V}(0) - \frac{1}{2}\int\frac{d^dk}{(2\pi)^d}\beta^2\bar{\rho}V^2(k)\tilde{S}(k) + \ldots
\ee
which gives
\be
\frac{\partial W}{\partial \bar{\rho}} = \frac{W}{\bar{\rho}} - \frac{1}{2}\int\frac{d^dk}{(2\pi)^d}\beta^2\bar{\rho}V^2(k)\frac{\partial \tilde{S}(k)}{\partial \bar{\rho}} + \ldots.
\ee
Recalling that
\be
\tilde{S}(k) = \frac{1}{1-\bar{\rho}c(k)},
\ee
we have
\be
\frac{\partial \tilde{S}(k)}{\partial \bar{\rho}}
= \tilde{S}^2(k)\frac{\partial (\bar{\rho}c(k))}{\partial \bar{\rho}}
\ee
which finally gives
\be
\frac{\partial (\beta P)}{\partial \bar{\rho}}
&=&1+\bar{\rho}\beta V(0) - \frac{1}{2}\int\frac{d^dk}{(2\pi)^d}\beta^2V^2(k)
\bigg[S(k) + S^2(k)\frac{\partial (\bar{\rho}c(k))}{\partial \bar{\rho}}\bigg].
\ee
or, integrating,
\be
\beta P &=& \bar{\rho}+\frac{1}{2}\bar{\rho}^2\beta V(0) - \frac{1}{2}\int\frac{d^dk}{(2\pi)^d}\beta^2V^2(k)
\bigg[\bar{\rho}S(k) + \bar{\rho}S^2(k)c(k)\bigg]
\nonumber\\
&=& \bar{\rho}+\frac{1}{2}\bar{\rho}^2\beta V(0) - \frac{\beta^2}{2}\int\frac{d^dk}{(2\pi)^d}V^2(k)S^2(k).
\ee

\subsection{Comparison with Carnahan-Starling Form}

We can go one step further and compare our results with the Carnahan-Starling equation of state. This form is an approximate, but quite accurate, equation of state valid for hard spheres.

Recalling the Carnahan-Starling form\cite{CS-EOS, Hansen}, we have
\be
\frac{\beta P}{\bar{\rho}} = \frac{1+\eta+\eta^2-\eta^3}{(1-\eta)^3}.
\ee

Let us begin by taking the derivative of this with respect to $\bar{\rho}$,
\be
\frac{\partial(\beta P)}{\partial\bar{\rho}}
&=& \frac{\partial}{\partial \bar{\rho}}\bigg(\bar{\rho}\frac{1+\eta+\eta^2-\eta^3}{(1-\eta)^3}\bigg)
\nonumber\\
&=& \frac{1+\eta+\eta^2-\eta^3}{(1-\eta)^3}
+ \bar{\rho}\frac{\partial}{\partial \eta}\bigg(\frac{1+\eta+\eta^2-\eta^3}{(1-\eta)^3}\bigg)\frac{\partial \eta}{\partial \bar{\rho}}
\nonumber\\
&=& \frac{1+\eta+\eta^2-\eta^3}{(1-\eta)^3}
+ \eta\bigg(\frac{(1-\eta)(1+2\eta-3\eta^2)+3(1+\eta+\eta^2-\eta^3)}{(1-\eta)^4}\bigg)
\nonumber\\
&=& \frac{1+4\eta+4\eta^2-4\eta^3+\eta^4}{(1-\eta)^4}.
\ee
This result is the left-hand side of the thermodynamic relation in Eq. (\ref{eq:A1}). Let us now rewrite the right-hand side.

We have
\be
\frac{\bar{\rho}}{\rho_0}\frac{\partial \rho_0}{\partial \bar{\rho}}
&=& \frac{\bar{\rho}}{\rho_0}\bigg[\bar{\rho}e^{W(\eta)}\frac{\partial W(\eta)}{\partial \eta}\frac{\partial \eta}{\partial \bar{\rho}}+e^{W(\eta)}\bigg]
\nonumber\\
&=&\eta\frac{\partial W(\eta)}{\partial \eta} + 1
\ee
where we again use the general form for $\rho_0$ given by Eq. (\ref{eq:E10}).

Setting the left and right halves equal, we have
\be
\frac{1+4\eta+4\eta^2-4\eta^3+\eta^4}{(1-\eta)^4}
= \eta\frac{\partial W(\eta)}{\partial \eta} + 1.
\ee
which we may rearrange as
\be
\frac{\partial W(\eta)}{\partial \eta} = \frac{1+4\eta+4\eta^2-4\eta^3+\eta^4-(1-\eta)^4}{\eta(1-\eta)^4}
= \frac{2(4-\eta)}{(1-\eta)^4}.
\ee
Integrating, we find
\be
W(\eta) = \int_0^{\eta} dx \frac{2(4-x)}{(1-x)^4} = \frac{8\eta-9\eta^2+3\eta^3}{(1-\eta)^3}
\ee
or, returning to the full form for $\rho_0$,
\be
\rho_0 = \bar{\rho}\exp\bigg[\frac{8\eta-9\eta^2+3\eta^3}{(1-\eta)^3}\bigg].
\ee

We now have an independent measure for the quality of our equation of state results. As one self-consistently solves for the pseudo-potential, we may compare our perturbative result Eq.(\ref{eq:E1}) to this result.


\begin{thebibliography}{99}

\bibitem{FTSPD}
G. F. Mazenko, Phys. Rev. E {\bf 81}, 061102 (2010).
(Referred to throughout as FTSPD.)

\bibitem{SDENE}
G. F. Mazenko, Phys. Rev. E {\bf 83}, 041125 (2011).
(Referred to throughout as SDENE.)

\bibitem{SD}
Smoluchowski dynamics has become identified with the overdamped kinetics in colloidal systems where the momenta become equilibrated much faster than the positions and one has a dynamics which is subsequently organized in terms of  the positions.

Historical papers include the following: A. Einstein, Ann. d. Physik {\bf 17}. 549 (1905); M. V. Smoluchowski, Phys. Zeit. {\bf 17}, 557 (1916); and P. Langevin, Comptes. rendus {\bf 146}, 530 (1908).

Important papers treating the many-particle Smoluchowski dynamics system include the following: B. J. Ackerson, JCP {\bf 64}, 242 (1976) and W. Dietrich and I. Peschel, Physica A{\bf 95}, 208 (1979).

\bibitem{Newtonian}
S. P. Das and G. F. Mazenko, arXiv:1111.0571v1 (2011).

\bibitem{self-const}
By self-consistent, we mean that kernels in the treatment of the two- and three-particle kinetics are functionals of the exact two-point propagators.

\bibitem{KE_Dave}
D. D. McCowan, G. F. Mazenko, (unpublished).

\bibitem{KE_Paul}
P. Spyridis, G. F. Mazenko, (unpublished).

\bibitem{Crisanti}
A. Crisanti,  Nuc. Phys. B {\bf 796}, 425 (2008).

\bibitem{Goetze}
W. Goetze  in {\bf Liquids, Freezing and Glass Transition}, edited by J. P. Hansen, D. Levesque, and J.Zinn-Justin (North-Holland, Amsterdam, 1991) Chap. 5.

\bibitem{Das}
S. Das, {\bf Statistical Physics of Liquids at Freezing and Beyond}, (Cambridge University Press, New York, 2011); Rev. Mod. Phys. {\bf 76}, 785 (2004).

\bibitem{ABB}
A. Andreanov, G. Biroli and J.-P. Bouchaud, EPL {\bf 88}, 16001 (2009).

\bibitem{Leibowitz}
H. L. Frisch and J. L. Leibowitz, {\bf The Equilibrium Theory of Classical Fluids}, (Benjamin, New York, 1964).

\bibitem{cumvert}
G. F. Mazenko, D. D. McCowan, P. Spyridis, (unpublished).

\bibitem{MemFunc}
The original memory function method was develped in terms of a projection operator formalism: R. W. Zwanzig, in {\bf Lectures in Theoretical Physics, Vol. 3} (Interscience: New York, 1961); R. W. Zwanzig, Phys. Rev. {\bf 24}, 983 (1965); and H. Mori, Prog. Theor. Phys. (Kyoto) {\bf 33}, 423 (1965).

For a slightly different approach, see
G. F. Mazenko, Phys. Rev. A {\bf 7}, 209 (1973) and Phys. Rev. A {\bf 9}, 360 (1974).

\bibitem{SigmaBB}
In SDENE Eq.(83), a sign error led to an incorrect form reported as
\be
\rho_0\beta^2\Sigma_{BB}(q,t=0) = \rho_0S^{-1}(q)-[1+\rho\beta V(0)].
\ee
The division of the vertices is more clearly defined here and includes terms of higher order not addressed in SDENE, but the correct result at that level should have read
\be
\rho_0\beta^2\Sigma_{BB}(q,t=0) = [1+\rho\beta V(0)]-\rho_0S^{-1}(q).
\ee

\bibitem{CS-EOS}
N. F. Carnahan and K. E. Starling, J. Chem. Phys. {\bf 51}, 635 (1969).

\bibitem{Hansen}
J-P. Hansen and I.R. McDonald, {\bf Theory of Simple Liquids, Third Edition}, (Academic Press New York, 2006) Chap. 4.

\end{thebibliography}
\end{document}